\documentclass{rspublic}
\usepackage{amssymb,amsmath}

\begin{document}

\title[Integrability and Linearization]{On the complete integrability and 
linearization of certain second order nonlinear ordinary differential equations} 

\author[Chandrasekar, Senthilvelan and Lakshmanan]{V. K. Chandrasekar, 
M. Senthilvelan and M. Lakshmanan}

\affiliation{Centre for Nonlinear Dynamics, Department of Physics, 
Bharathidasan Univeristy, Tiruchirapalli - 620 024, India}

\label{firstpage}

\maketitle

\begin{abstract}{Integrability, Integrating factor, Linearization, Equivalence
problem}
A method of finding general solutions of second-order  nonlinear 
ordinary differential 
equations by extending the Prelle-Singer (PS) method is briefly discussed.  
We explore integrating factors, integrals of motion and the general
solution associated with several dynamical systems discussed in the current
literature by employing our modifications and extensions of the 
PS method.  In addition to the above we introduce a novel way of deriving 
linearizing transformations
from the first integrals to linearize the second order nonlinear ordinary 
differential equations to free particle equation.  We illustrate the theory 
with several potentially important examples and show that our procedure is
widely applicable.

\end{abstract}
                                                                               
\section{Introduction}
\label{sec1}
Solving nonlinear ordinary differential equations (ODEs) is one of the
classical but potentially important areas of research in the theory of 
dynamical systems (Arnold 1978; Jose \& Saletan 2002; Wiggins 2003).  
Indeed the last century has witnessed considerable amount of 
research activity in this field.  Progress has been made broadly in two 
different ways: one is through geometrical analysis and the other is 
through analytical studies.    
The modern geometrical theory was born with Poincar\'e and developed 
later vigorously by Arnold, Moser, Birkhoff and others 
(Percival \& Richards 1982; Guckenheimer \& Holmes 1983; Wiggins 2003).  
In parallel with the geometrical theories various analytical methods have also 
been devised to tackle nonlinear ODEs.  The ideas developed by Kovalevskaya,  
Painlev\'e and his coworkers have been used 
to integrate a class of nonlinear ODEs and obtain their underlying solutions 
(Ince 1956).  As a consequence of 
these studies, nonlinear dynamical systems are broadly classified
into two categories, namely, (i) integrable and (ii) nonintegrable systems. 
Indeed one of the important current problems in nonlinear dynamics is to 
identify integrable dynamical systems (Lakshmanan \& Rajasekar 2003).  
Of course,
these methods have close connection with the group theoretical approach
introduced by Sophus Lie in the nineteenth century and subsequently extended by
Cartan and Tresse, to integrate ordinary and partial differential equations (see
for example Olver (1995); Bluman \& Anco (2002)).

In order to identify such integrable dynamical systems different techniques 
have been proposed, including Painlev\'e analysis (Conte 1999), Lie 
symmetry analysis (Bluman \& Anco 2002) and direct methods of finding 
involutive integrals of motion (Hietarinta 1987). 
Each method has its own advantages and disadvantages.  For a detailed discussion
about the underlying theory of each method, their limitations and  
applications we refer to Lakshmanan \& Rajasekar (2003).  Also certain 
nonlinear ODEs  
can be solved by transforming them to linear ODEs whose solutions are known. 
In fact, linearization of given nonlinear ODEs is one of the 
classical problems in the theory of ODEs
whose origin dates back  to Cartan.  For the recent progress in this direction 
we refer Olver (1995).
  
In this direction sometime ago Prelle \& Singer (1983) have proposed a 
procedure for solving first order ODEs that presents the solution in terms 
of elementary functions if such a solution exists.  The attractiveness 
of the PS method is that if the given system of first order ODEs has a solution 
in terms of elementary functions then the method guarantees that this solution 
will be found.  Very recently Duarte \textit{et al.} (2001) modified the 
technique 
developed by Prelle \& Singer (1983) and applied it to second order ODEs.   
Their approach was based on the conjecture that if an elementary solution 
exists for the given second order ODE then there exists at least one 
elementary first integral $I(t,x,\dot{x})$ whose derivatives are all 
rational functions of $t$, $x$ and $\dot{x}$.  For a class of  
systems these authors (Duarte \textit{et al.} 2001)  have deduced first 
integrals and in some cases for the first time through their procedure.  

In this paper we show that the theory of Duarte \textit{et al.} (2001) can be 
extended in different directions to isolate even two independent integrals of 
motion and obtain solutions explicitly.  In the earlier study
it has been shown that the theory can be used to derive only one integral.  
In this work we extend their theory and deduce general solution from the 
first integral.  Our examples include those considered in their paper and 
certain important  
equations discussed in the recent literature whose solutions are not known.   
Our study was motivated by two reasons.  Firstly, it is to show that apart from 
finding first integrals one can also deduce general solutions in a 
straightforward and simple manner.  Here the method  
we propose is not confined to the PS method alone but can be treated as a 
\emph{general one}, that is, suppose one has a first integral 
for a given second order ODE then our method provides the general solution 
in an algorithmic way at least for a class of equations. The 
reason for merging our procedure with PS method rather 
than any other method is due to the following facts.  (1) 
For a given problem if the solution exists it has been conjectured 
that the PS method guarantees to provide first integrals.  (2) The PS method 
not only gives the first 
integrals but also the underlying integrating factors, that is, multiplying 
the equation with these functions we can rewrite the equation as a  perfect 
differentiable function which upon integration gives first integrals in a 
separate way.  (3) The PS method can be used 
to solve nonlinear as well as linear second order ODEs.  As the PS method 
is based on the equations of motion rather than Lagrangian or Hamiltonian 
the analysis is applicable to deal with both Hamiltonian and non-Hamiltonian 
systems.  

Our second reason is to bring out a novel and straightforward way to 
construct linearizing transformations.  Particularly we demonstrate that using 
our procedure one can also explore linearizing transformations in a simple 
and straightforward manner.  The latter can be used to transform the 
given second order nonlinear ODEs to a linear one, in particular, to 
the free particle equation.  As we illustrate below these transformations 
can be deduced from the first integral itself which is totally 
\emph{a new technique} in the current literature.  In a nutshell,  
once a first integral is known then our procedure, at least for a class 
of problems, not only gives the general solutions but also 
provides linearizing transformations.  The ideas proposed here can be applied 
to coupled system of second order ODEs as well as higher order ODEs, which 
will be presented separately.   

	The paper is organized as follows.  In the following section
we briefly describe the Prelle-Singer method applicable for second order ODEs
and indicate certain new features in finding the integrals of motion. 
In \S3, we have extended the theory in three different directions which
indicates the novelty of the approach. The first significant application is that
the second integral can be deduced straightforwardly from the method itself in
many cases. The
second one is to deduce the general solution from the first integral. Finally we
propose a method to identify linearizing transformations. 
We emphasize the validity of the theory with several illustrative examples arising 
in different areas of physics in \S4.  In \S5, we demonstrate 
the method of identifying linearizing transformation with three examples 
including the one studied in the recent literature. We present 
our conclusions in \S6.
	
\section{Prelle-Singer method for second order ODEs}
\label{sec2}
In this section, we briefly discuss the theory introduced by 
Duarte \textit{et al.} (2001) for second order ODEs and extend it suitably such 
that the general solutions can be deduced from the modifications.  Let us 
consider the second order ODEs of the form 
\begin{align} 
\ddot{x}=\frac{P}{Q},  \quad 
{P,Q}\in \mathbb{C}{[t,x,\dot{x}]}, 
 \label{met1}
\end{align}
where over dot denotes differentiation with respect to time and $P$ and $Q$ 
are polynomials in $t$, $x$ and $\dot{x}$ with coefficients in the 
field of complex numbers.  Let us assume that the ODE (\ref{met1}) 
admits a first integral $I(t,x,\dot{x})=C,$ with $C$ constant on the 
solutions, so that the total differential gives
\begin{eqnarray}  
dI={I_t}{dt}+{I_{x}}{dx}+{I_{\dot{x}}{d\dot{x}}}=0, 
\label{met3}  
\end{eqnarray}
where the subscript denotes partial differentiation with respect 
to that variable.  Rewriting equation~(\ref{met1}) in the form 
$\frac{P}{Q}dt-d\dot{x}=0$ and adding a null term 
$S(t,x,\dot{x})\dot{x}$ $ dt - S(t,x,\dot{x})dx$ to the latter, we obtain that on 
the solutions the 1-form
\begin{eqnarray}
\bigg(\frac{P}{Q}+S\dot{x}\bigg) dt-Sdx-d\dot{x} = 0. 
\label{met6} 
\end{eqnarray}	
Hence, on the solutions, the 1-forms (\ref{met3}) and 
(\ref{met6}) must be proportional.  Multiplying (\ref{met6}) by the 
factor $ R(t,x,\dot{x})$ which acts as the integrating factors
for (\ref{met6}), we have on the solutions that 
\begin{eqnarray} 
dI=R(\phi+S\dot{x})dt-RSdx-Rd\dot{x}=0, 
\label{met7}
\end{eqnarray}
where $ \phi\equiv {P}/{Q}$.  Comparing equations (\ref{met3}) 
with (\ref{met7}) we have, on the solutions, the relations 
\begin{align} 
 I_{t} & = R(\phi+\dot{x}S), \nonumber\\
 I_{x} & = -RS, \nonumber\\
 I_{\dot{x}} & = -R.  
 \label{met8}
\end{align} 
Then the compatibility conditions, 
$I_{tx}=I_{xt}$, $I_{t\dot{x}}=I_{{\dot{x}}t}$,
$I_{x{\dot{x}}}=I_{{\dot{x}}x}$, between the equations (\ref{met8}), require
that 
\begin{align}  
D[S] & = -\phi_x+S\phi_{\dot{x}}+S^2,\label{met9}\\
D[R] & = -R(S+\phi_{\dot{x}}),\label{met10}\\
R_x & = R_{\dot{x}}S+RS_{\dot{x}},
\label{met11}
\end{align}
where 
\begin{eqnarray}
D=\frac{\partial}{\partial{t}}+
\dot{x}\frac{\partial}{\partial{x}}+\phi\frac{\partial}
{\partial{\dot{x}}}.
\nonumber
\end{eqnarray}

Equations~(\ref{met9})-(\ref{met11}) can be solved in the following way.  
Substituting the given expression of $\phi$ into (\ref{met9}) and solving 
it one can obtain an expression for $S$.  Once $S$ is known then  
equation~(\ref{met10}) becomes the determining equation for the function $R$.  
Solving the latter one can get an explicit form for $R$.  
Now the functions $R$ and $S$ have to satisfy an extra constraint, that is, 
equation~(\ref{met11}).  Once a compatible solution satisfying all the three 
equations have 
been found then the functions $R$ and $S$ fix the integral of motion 
$I(t,x,\dot{x})$ by the relation 
\begin{align}
I(t,x,\dot{x}) & = \int R(\phi+\dot{x}S)dt
  -\int \left( RS+\frac{d}{dx}\int R(\phi+\dot{x}S)dt \right) dx
  \notag\\
  & -\int \left\{R+\frac{d}{d\dot{x}} \left[\int R (\phi+\dot{x}S)dt-
   \int \left(RS+\frac{d}{dx}\int R(\phi+\dot{x}S)dt\right)dx\right] 
  \right\}d\dot{x}.
  \label{met13}
\end{align}
Equation~(\ref{met13}) can be derived straightforwardly by 
integrating the equations~(\ref{met8}).
Note that for every independent set $(S,R)$, equation~(\ref{met13}) defines an
integral.

Thus two independent sets, $(S_i,R_i),\;i=1,2,$ provide us two independent
integrals of motion through the relation (\ref{met13}) which guarantees the
integrability of equation~(\ref{met1}). We noticed that since we are solving
equations~(\ref{met9})-(\ref{met10}) first and check the compatibility of this
solution with equation (\ref{met11}), one often meets the situation that 
all the solutions which satisfy 
equations~(\ref{met9})-(\ref{met10}) need not satisfy the constraint 
(\ref{met11}) 
since equations~(\ref{met9})-(\ref{met11}) constitute an overdetermined system 
for the unknowns $R$ and $S$. In fact, for a class of problems one often gets a
set $(S_1,R_1)$ which satisfies all the three equations~(\ref{met9})-(\ref{met11})
and another set $(S_2,R_2)$ which satisfies only the first two equations and not
the third, namely, (\ref{met11}). In this situation, we find the interesting
fact that one can use the first integral, derived from the set $(S_1,R_1)$, 
and deduce the second compatible solution $(S_2,\hat{R_2})$.  For example, let the 
set $(S_2,R_2)$ be a 
solution of equations~(\ref{met9})-(\ref{met10}) and not of the constraint 
equation~(\ref{met11}).  After examining several examples we find that 
one can make 
the set $(S_2,R_2)$ compatible by modifying the form of $R_2$ as
\begin{align}
\hat{R_2} & = F(t,x,\dot{x})R_2,
\label{met101}
\end{align} 
where $\hat{R_2}$ satisfies equation~(\ref{met10}), so that we have
\begin{eqnarray} 
(F_t+\dot{x}F_x+\phi F_{\dot{x}}) R_2+FD[R_2]
   =-FR_2(S_2+\phi_{\dot{x}}).
\label{met102}
\end{eqnarray}
Further, if $F$ is a constant of motion (or a function of it), then the first 
term on the left hand side
vanishes and one gets the same equation~(\ref{met10}) for $R_2$ provided $F$ is 
non-zero. In other words, whenever $F$ is a constant of motion or a function 
of it then the solution of equation~(\ref{met10}) may provide only a factor 
of the complete solution $\hat{R_2}$ without the factor $F$ in 
equations~(\ref{met101}).  This general form of $\hat{R_2}$ with $S_2$ 
can now form a complete
solution to the equations~(\ref{met9})-(\ref{met11}).  In a nutshell we 
describe the procedure as follows.  First we determine $S$ and $R$ from 
equations~(\ref{met9})-(\ref{met10}). If the set $(S,R)$ satisfies 
equation~(\ref{met11})
then we take it as a compatible solution and proceed to construct the associated
integral of motion.  On the other hand if it does not
satisfy (\ref{met11}) then we assume the modified form $\hat{R_2}=F(I_1)R_2$, 
where $I_1$
is the first integral which has already been derived through a compatible
solution, and 
find the explicit form of $F(I_1)$  from equation~(\ref{met11}), which in turn 
fixes the compatible solution $(S_2,\hat{R_2})$.  This set $(S_2, \hat{R_2})$ can be
utilized to derive the second integral. 
\section{Generalization}
\label{sec3}
\subsection{Identifying a second integral of motion}
\label{sec3.1}
Duarte \textit{et al.} (2001) have considered certain physically important systems 
and constructed first integrals.  Further, they mentioned that applying the
original PS algorithm
to these first integrals (by treating them as first order ODEs) one can 
deduce the general solution.  An interesting observation we make here is 
that there is no need to invoke the
original PS procedure in order to deduce the general solution. In fact, as we
show below, the general solution can be derived in a self-contained way. 
As the motivation of Duarte \textit{et al.} (2001) was to construct only
the first integral they reported only one set of solution $(S,R)$ for the 
equations~(\ref{met9})-(\ref{met11}).  However, we have observed that an 
additional independent
set of solution, namely, $(S_2,R_2)$, of equations~(\ref{met9})-(\ref{met11})
 may lead to another integral of motion, $I_2$, and 
if the latter is an independent function of $I_1$ then one can write down the 
general solution for the given problem from these two integrals alone
straightforwardly.  Now the question is 
whether one will be able to find a second pair of solution for the system 
(\ref{met9})-(\ref{met11}) and construct $I_2$ 
through the relation (\ref{met13}).  After investigating several 
examples we observed the following.  (i) For 
a class of equations, including that of harmonic oscillator, 
equation coming from general relativity and the generalized modified Emden 
equation
with constant external forcing one can construct a second pair 
of solution  $(S_2,R_2)$ and deduce $I_2$ through the relation (\ref{met13}) 
straightforwardly.  We call 
this class as Type I.  (ii) For another class of equations 
we can find explicitly $(S_2,R_2)$ from (\ref{met9})-(\ref{met11})
but one is unable to integrate equation~(\ref{met13}) exactly and obtain the 
second integral $I_2$ explicitly.  We call this class as Type II.  
The examples included in this category are 
Helmhotz oscillator and Duffing oscillator.  For this class 
of equations we 
identify an alternate way to derive the second integration constant. (iii) 
Besides 
the above there exists another category in which the systems do not even admit 
a second pair $(S_2,R_2)$ of solution in simple rational forms
for the equations~(\ref{met9})-(\ref{met11}) 
and we call this  category as Type III.  An example 
under this class is the Duffing - van der Pol oscillator which is one of the 
prototype examples for the study of nonlinear dynamics in many branches of 
science.  For this class of equations also
we identify an alternate way to obtain the second integral. 
\subsection{Method of deriving general solution}
\label{sec3.2}
To overcome the difficulties in constructing the second constant in 
Types II and III  we propose the following procedure.
As our aim is to derive the general solution for the given problem, we 
split the functional form of the first integral $I$ into two terms such that 
one involves all the variables $(t,x,\dot{x})$ while the other excludes 
$\dot{x}$, that is, 
\begin{eqnarray} 
I=F_1(t,x,\dot{x})+F_2(t,x). 
\label{met13a}
\end{eqnarray}
Now  let us split the function $F_1$ further in terms of two functions such 
that $F_1$ itself is a function of the product of the two functions, say, a 
perfect differentiable function $\frac {d}{dt}G_1(t,x)$ and another function 
$G_2(t,x,\dot{x})$, that is,
\begin{eqnarray} 
I=F_1\left(\frac{1}{G_2(t,x,\dot{x})}\frac{d}{dt}G_1(t,x)\right)
+F_2\left(G_1(t,x)\right).
\label{met13b}
\end{eqnarray}
We note that while rewriting equation (\ref{met13a}) in the form 
(\ref{met13b}), we demand the function $F_2(t,x)$ in (\ref{met13a}) 
automatically to be a function of $G_1(t,x)$. The reason for making such a
specific decomposition is that in this case one can rewrite
equation~(\ref{met13b}) as a simple first order ODE for the variable $G_1$ 
(see equation~(\ref{met13d}) below). Actually, we realized originally such a
possibility for the case of the integrable force-free Duffing-van der Pol
oscillator equation (Chandrasekar \textit{et al.} 2004), which has now been 
generalized in the present case. Now identifying  the function  
$G_1$ as the new dependent variable and the integral of $G_2$ over time as 
the new independent variable, that is,
\begin{align} 
w = G_1(t,x),\quad z = \int_o^t G_2(t',x,\dot{x}) dt', 
\label{met13c}
\end{align}
one indeed obtains an explicit transformation to remove  the time dependent 
part in the first integral (\ref{met13}).  We note here that the integration 
on the right hand side of (\ref{met13c}) leading to $z$ can be performed 
provided the function $G_2$ is an exact derivative of $t$, that is,
$G_2=\frac{d}{dt}z(t,x)=\dot{x}z_x+z_t$, so that $z$ turns 
out to be a function
$t$ and $x$ alone.  In terms of the new variables, equation (\ref{met13b}) can 
be modified to the form
\begin{eqnarray} 
I=F_1\left(\frac {dw}{dz}\right)+F_2(w).
\label{met13d}
\end{eqnarray}
In other words,
\begin{eqnarray} 
F_1\left(\frac {dw}{dz}\right)=I-F_2(w).
\label{met13e}
\end{eqnarray}
Now rewriting equation (\ref{met13d}) one obtains a separable equation 
\begin{eqnarray}  
\frac {dw}{dz}=f(w),
\end{eqnarray}
which can lead to the solution after an integration.  Now rewriting the 
solution in terms 
of the original variables one obtains a general solution for (\ref{met1}).
\subsection{Method of deriving linearizing transformations}
\label{sec3.3}
Finally, one may note the following interesting point in the above analysis.
Suppose $F_2(w)$ is zero in equation~(\ref{met13d}). Then one obtains the simple
equation 
\begin{eqnarray}  
\frac {dw}{dz}=\hat{I},
\end{eqnarray}
where $\hat{I}$ is a constant.  In other words we have
\begin{eqnarray}  
\frac {d^2w}{dz^2}=0,
\end{eqnarray} 
which is nothing but the free particle equation.  In  this case the new
variables $z$ and $w$ helps us to transform the given second order nonlinear
ODE into a
linear second order ODE which in turn leads to the solution by trivial
integration.  The new variables $z$ and $w$ turn out to be the linearizing
transformations.  We discuss in detail about this possibility separately in \S5.
\section{Applications}
\label{sec4}
In this section, we demonstrate the theory discussed in the previous section
with suitable examples.    
In particular, we consider several interesting examples, including 
those considered in Duarte \textit{et al.} (2001) and derive general solutions 
and establish 
complete integrability of these dynamical systems.  We split 
our analysis into three categories.  In the first category we consider 
examples in which the $I_i's,\;i=1,2$, can be derived straightforwardly 
from the relation (\ref{met13}).  
In the second and third categories we follow our own procedure detailed in
\S3$\,b\,$ and \S3$\,c\,$ and deduce 
the second constant.  We note that one can apply our procedure to a wide 
range of systems 
of second order of type (\ref{met1}) but for illustration purpose we 
consider only a few examples in the following.    
\subsection{Type - I Systems}
\label{sec4.1}
As mentioned earlier for certain equations one can also get the second 
pair of solution $(S_2,R_2)$, 
in an algorithmic way, from the determining equations (\ref{met9})-(\ref{met11}) 
and construct $I_2$ through the relation (\ref{met13}).  We observe that 
the Examples 1 and 2 discussed in Duarte \textit{et al.} (2001) can be solved 
in this way and so we 
consider these two examples first and then a nontrivial example 
in the following.   
\subsubsection*{Example 1 : An exact solution in general relativity}
Duarte \textit{et al.} (2001) considered the following equation, which was 
originally derived by Buchdahl (1967) in the theory of general relativity,
\begin{eqnarray}            
x\ddot{x}=3\dot{x}^2+\frac{x\dot{x}}{t},
\label{cat11}
\end{eqnarray}
and deduced the first integral $I$ through their procedure.  In the following, 
we briefly discuss their results and then illustrate our ideas.  
Substituting $\phi = \frac{3\dot{x}^2}{x}+\frac{\dot{x}}{t}$ into
(\ref{met9})-(\ref{met11}) we get 
\begin{align}           
S_t+\dot{x}S_x+\frac{\dot{x}(3t\dot{x}+x)}{tx}S_{\dot{x}}
 & = \frac{3\dot{x}^2}{x^2}+\frac{6t\dot{x}+x}{tx}S+S^2,
 \label{cat12}\\
R_t+\dot{x}R_x+\frac{\dot{x}(3t\dot{x}+x)}{tx}R_{\dot{x}}
& = -RS-\frac{6\dot{x}t+x}{tx}R,
\label{cat13}\\
R_x  & = SR_{\dot{x}}+RS_{\dot{x}}.
\label{cat14}
\end{align} 
As mentioned in Sec.~2 let us first solve equation~(\ref{cat12}) and obtain 
an explicit form for the function $S$.  
To do so Duarte \textit{et al.} (2001) consider an ansatz for $S$ of the form 
\begin{eqnarray} 
S = \frac{a(t,x)+b(t,x)\dot{x}}{c(t,x)+d(t,x)\dot{x}},
\label{cat15}
\end{eqnarray}
where $a$, $b$, $c$ and $d$ are arbitrary functions of $t$ and $x$.  
A rational form for $S$ can be justified, since from (\ref{met8}) one may 
note that $S=\frac{I_x}{I_{\dot{x}}}$.  So we consider only
rational forms in $\dot{x}$ for $S$ for all the examples which we consider in 
this paper. One may note that in certain examples, including the present one and
examples 3 and 5 (given below), this form degenerates into a polynomial form in
$\dot{x}$. However, for other examples like the examples 2 and 4 below, one
requires a rational form such as (\ref{cat15}). To be general, we carry out an
analysis with the form (\ref{cat15}).

Substituting (\ref{cat15}) into (\ref{cat12}) and equating the coefficients 
of different powers of $\dot{x}$ to zero we get a set of partial differential 
equations for the variables $a$, $b$, $c$ and $d$.  Solving them one obtains
\begin{equation}
S_1 = -\frac{3\dot{x}}{x}, \quad S_2=-\frac{\dot{x}}{x}.
\label{new1}
\end{equation}

We note that Duarte \textit{et al.} (2001) have reported only the expression 
$S_1$, as the
solution for equation~(\ref{cat12}).  However, we find $S_2$ also forms a 
solution for (\ref{cat12}) and helps to deduce the general solution. 
Substituting these forms $S_1$ and $S_2$ into (\ref{cat13}) and solving the 
latter one can get an explicit form for the function $R$.  Let us first 
consider $S_1$.  Substituting the latter into (\ref{cat13}) we get the 
following equation for $R$:
\begin{equation}
R_t+\dot{x}R_x+\frac{\dot{x}(3t\dot{x}+x)}{tx}R_{\dot{x}}
 = \frac{3\dot{x}}{x}R-\frac{6\dot{x}t+x}{tx}R.
\label{new2}
\end{equation}
In order to solve (\ref{new2}) again one has to make an ansatz.  We assume 
that the following form for $R$, that is,
\begin{equation}
R = A(t,x)+B(t,x)\dot{x},
\label{new3}
\end{equation}
where $A$ and $B$ are arbitrary functions of $(t,x)$.  Since $R=-I_{\dot{x}}$
(vide equation~(\ref{met8})) the form of $R$ may be a polynomial or rational 
in $\dot{x}$.  
So depending upon the problem one has to choose an appropriate ansatz. To begin 
with one can consider simple polynomial (in $\dot{x}$) for $R$ and if it fails 
then one can go for rational forms.  Let us start with equation~(\ref{new3}). 
Now substituting 
(\ref{new3}) into (\ref{new2}) and equating the coefficients of different 
powers of $\dot{x}$ to zero and solving the resultant equations one can obtain 
finally $R_1 = \frac{1}{tx^3}$.  Now the solution $S_1 = -\frac{3\dot{x}}{x}$ 
and $R_1 = \frac{1}{tx^3}$ has to satisfy the equation~(\ref{cat14}) in order to 
be a compatible solution, which is indeed true.  Once $R$ and $S$ have been 
found the first integral $I$ can be fixed easily using the expression 
(\ref{met13}) as 
\begin{equation}
I_1 = \frac{\dot{x}}{tx^3}.
\label{cat16}
\end{equation} 
One can easily check that $I_1$ is constant on the solutions, that is, 
$\frac{dI_1}{dt} = 0$.  This integral 
has been deduced in Duarte \textit{et al.} (2001).  However, the second expression, 
$S_2$ has been ignored by the authors since the corresponding $R_2$ coming 
out of (\ref{cat13}) does not form a compatible 
solution, that is, it does not satisfy (\ref{cat14}).  But in the following we 
show how one can make it compatible and use it effectively to deduce the 
second integration constant.  

Now substituting the expression $S_2=-\frac{\dot{x}}{x}$ into (\ref{cat13}) 
and solving it in the same way as outlined in the previous paragraph we 
obtain the following form for $R$, namely, 
\begin{equation} 
R_{2}=\frac{1}{x^5t}.
\label{new4}
\end{equation}
However, this set $(S_2,R_2)$ does not satisfy the extra constraint 
(\ref{cat14}).  In fact, not all forms of $R$ coming out from (\ref{met10}) 
satisfy (\ref{met11}) in general.  As we explained in \S3, the form of
$R_2$ given in (\ref{new4}) may not be the 'complete form' but might be a factor of
the complete form.  To recover the complete form of $R$ one may assume that 
\begin{equation}
\hat{R} = F(I_1)R,
\label{new6}
\end{equation}
where $F(I_1)$ is a function of the first integral $I_1$, and determine the form
of $F(I_1)$ explicitly. For this purpose we proceed as follows.  Substituting 
\begin{align}
\hat{R}_2 & = F(I_1)R_2 = \frac{1}{tx^5}F(I_1)
\label{cat17}
\end{align}
into equation(\ref{cat14}), we obtain an equation for $F$, that is, 
\begin{equation}
I_1F'+2F = 0,
\label{new7}
\end{equation}
where prime denotes differentiation with respect to $I_1$.  
Upon integrating (\ref{new7}) we get (after putting the constant of integration
to zero) 
\begin{eqnarray}
F = \frac {1}{I_1^2} = \frac{t^2x^6}{\dot{x}^2},
\label{cat18}
\end{eqnarray} 
which fixes the form of $\hat{R_2}$ as 
\begin{eqnarray}         
\hat{R}_2 = \frac {1}{I_1^2}\frac{1}{x^5t} = \frac{tx}{\dot{x}^2}. 
\label{cat19}  
\end{eqnarray}
Now one can easily check that this set $S_2 = -\frac{\dot{x}}{x}$ and 
$\hat{R}_2 = \frac{tx}{\dot{x}^2}$ is a compatible solution for the 
equations~(\ref{cat12})-(\ref{cat14}).  Substituting $S_2$ and $\hat{R}_2$ into
(\ref{met13}) we can get an explicit form for $I_2$, namely,
\begin{eqnarray} 
I_2 = t(t+\frac{x}{\dot{x}}).
\label{cat100}
\end{eqnarray}
From the integrals $I_1$ and $I_2$ one can deduce the general solution 
directly (without performing any further integration) for the problem in the 
form 
\begin{eqnarray} 
x=\sqrt\frac {1} {I_1(I_2-t^2)} .
\label{app38} 
\end{eqnarray}
Of course the same result can be obtained solving equation~(\ref{cat16}) from the 
first
integral.  However, the point we want to emphasize here is that an independent
second integral of motion can be deduced to find the solution without any
further integration, which can be used profitably when the expression for 
$I_1$ cannot be solved straightforwardly.

\subsubsection*{Example 2 : Simple harmonic oscillator}
To illustrate the above procedure also works for linear ODEs, in the 
following we consider the simple harmonic oscillator and derive
the general solution.  As the procedure of deriving the first integral 
has been discussed in detail in Duarte \textit{et al.} (2001) we omit the 
details and provide only the essential expressions in the following.  
 
The equation of motion for the simple harmonic oscillator is
\begin{eqnarray}            
\ddot{x}=-x
\label{cat101}
\end{eqnarray}
so that the equations~(\ref{met9})-(\ref{met11}) become 
\begin{align}           
S_t+\dot{x}S_x-x S_{\dot{x}} & = 1+S^2,
\label{cat102}\\
R_t+\dot{x}R_x-x R_{\dot{x}} & = -RS,
\label{cat103}\\
R_x-SR_{\dot{x}}-RS_{\dot{x}} & = 0.
\label{cat104}                                                         
\end{align} 
As shown in Duarte \textit{et al.} (2001) a simple solution for the 
equations~(\ref{cat102})-(\ref{cat104}) can be constructed of the form 
\begin{equation}  
S_1 = \frac{x}{\dot{x}},  \quad R_1 = \dot{x}
\label{cat105}
\end{equation}
which in turn gives the first integral 
\begin{eqnarray}     
I_1 =\dot{x}^2+ x^2,  
\label{cat106}
\end{eqnarray} 
through the relation (\ref{met13}).  However, one can easily check that 
\begin{equation}  
S_2=-\frac{\dot{x}}{x},\quad R_2=x
\label{cat107}
\end{equation}
is also a solution for the set  (\ref{cat102}) and (\ref{cat103}) (which has
not been reported earlier) but does not satisfy the extra constraint 
(\ref{cat104}).  So, as before, let us seek an $\hat{R}_2$ of the form
\begin{eqnarray}
\hat{R}_2 = F(I_1)R_2 = F(I_1)x,
\label{new211}
\end{eqnarray}
where $F(I_1)$ is a function of $I_1$.  
Substituting (\ref{new211}) into equation~(\ref{cat104}) and integrating the 
resultant equation, we get $F=\frac{1}{I_1}$.  Thus $\hat{R}_2$ becomes 
\begin{eqnarray}         
\hat{R}_2 = \frac{x}{I_1} = \frac{x}{x^2+\dot{x}^2}.  
\label{cat109} 
\end{eqnarray}
Now one can check that $(S_2,\hat{R}_2)$ satisfies all three equations 
(\ref{cat102})-(\ref{cat104}) and furnishes the second integral through 
the relation (\ref{met13}) of the form 
\begin{align}   
I_2 & = -t - \int \frac{\dot{x}}{\dot{x}^2+x^2} dx 
- \int {\bigg ( \frac{x}{\dot{x}^2+x^2} 
-\frac{d}{d\dot{x}} \int \frac{\dot{x}}
{\dot{x}^2+x^2}dx\bigg )} d\dot{x},
\nonumber\\
 & = -t -\tan ^{-1} \frac{\dot{x}}{x}.
\label{cat1001}
\end{align}
Using (\ref{cat106}) and (\ref{cat1001}) we can write down the general 
solution for the simple harmonic oscillator directly in the form
\begin{eqnarray}
x=\sqrt{I_1} \cos(t + I_2).
\label{cat1002}
\end{eqnarray}
In a similar way one can deduce general solution for a class of physically 
important systems.

One may note that in the above two examples, $I_2$ can also be obtained
trivially by simply integrating the expressions (\ref{cat16}) and 
(\ref{cat106})
without using the extended procedure.  We stress that for certain equations
one is not able to integrate and obtain the general solution  in this
simple way and has to follow the above said procedure in order to obtain the
second integral.  In the following we discuss one such example for which to our
knowledge explicit solution was not known earlier. 
\subsubsection*{Example 3 : Modified Emden type equation with linear term}
It is known that the generalized Emden type equation with linear and 
constant external forcing is also linearizable since it admits an eight point 
Lie symmetry group (Mahomed \& Leach 1989; Pandey \textit{et al.} 2004).  
In the following we explore its general solution through the extended PS 
algorithm.  Let us first consider the equation of the form
\begin{eqnarray}            
\ddot{x}+kx\dot{x}+\frac {{k}^2}{9}x^3+\lambda_1x=0,\label {lam101}
\end{eqnarray}
where $k$ and $\lambda_1$ are arbitrary parameters.  To explore the general
solution for the equation~(\ref{lam101}) we again use the PS method. In 
this case we have the  following determining equations for the functions 
$R$ and $S$,
\begin{align}            
S_t+\dot{x}S_x-(kx\dot{x}+\frac {k^2}{9}x^3+\lambda_1x) S_{\dot{x}}
& = k\dot{x}+\frac {k^2}{3}x^2+\lambda_1-S kx+S^2,\label {lam102}\\
R_t+\dot{x}R_x-(kx\dot{x}+\frac {k^2}{9}x^3+\lambda_1x) R_{\dot{x}}
& = -R(S-kx),\label {lam103}\\
R_x-SR_{\dot{x}}-RS_{\dot{x}} & = 0.\label {lam104}
\end{align}

As before, let us seek an ansatz for $S$ of the form (\ref{cat15})
to the first equation in (\ref{lam102})-(\ref{lam104}).  Substituting the 
ansatz (\ref{cat15}) into (\ref{lam102}) and equating the coefficients 
of different powers of $\dot{x}$ to zero we get
\begin{align}          
db_x-bd_x -kd^2 & = 0 ,\nonumber\\
db_t-bd_t+cb_x-bc_x+a_xd-ad_x 
-2kcd-(\frac{k^2}{3}x^2+\lambda_1)d^2+kbdx-b^2 & = 0,\nonumber\\
cb_t-bc_t+da_t-ad_t+ca_x-ac_x 
-kc^2-2(\frac{k^2}{3}x^2+\lambda_1)cd+2kadx-2ab & = 0,\nonumber\\
ca_t-ac_t-(\frac {k^2}{9}x^3+\lambda_1x)(bc-ad) 
-(\frac{k^2}{3}x^2+\lambda_1)c^2+kacx-a^2 & = 0,
\label{lam106}
\end{align}
where subscripts denote partial derivative with respect to that variable.
Solving equation~(\ref{lam106}) we can obtain two specific solutions,
\begin{equation}
S_1 =\frac{-\dot{x}+\frac{k}{3}x^2}{x}, \quad 
S_2=\frac{kx+3\sqrt{-\lambda_1}}{3}-\frac{k\dot{x}}
{kx+3\sqrt{-\lambda_1}}.
\label{lam107}
\end{equation}
Putting the forms of $S_1$ and $S_2$ into (\ref{lam103}) and solving it 
one can obtain the respective forms of $R$.  To do so let us first 
consider $S_1$.  Substituting the latter into (\ref{lam103}) we get the 
following equation for $R$:
\begin{equation}
R_t+\dot{x}R_x-(kx\dot{x}+\frac {{k}^2}{9}x^3+\lambda_1x) R_{\dot{x}}
=\bigg(\frac{\dot{x}-\frac{k}{3}x^2}{x}+kx \bigg)R.
\label{lam108}
\end{equation}

Again to solve equation~(\ref{lam108}) we make an ansatz of the form
\begin{equation}
R = \frac{A(t,x)+B(t,x)\dot{x}}{C(t,x)+D(t,x)\dot{x}+E(t,x)\dot{x}^2}.
\label{lam109}
\end{equation} 
Now substituting 
(\ref{lam109}) into (\ref{lam108}) and equating the coefficients of different 
powers of $\dot{x}$ to zero and solving the resultant equations we arrive at 
\begin{equation}
R_1= e^{-2\sqrt{-\lambda_1}t}\bigg( \frac{C_0x}{(3\dot{x}+kx^2-
3\sqrt{\lambda_1}x)^2}\bigg),
\label{lam110}
\end{equation}
where $C_0=18\sqrt{-\lambda_1}$.  One can easily check that $S_1$ and $R_1$ 
satisfies equation~(\ref{lam104}) and as a consequence one obtains the first 
integral 
\begin{eqnarray}
I_1=e^{-2\sqrt{-\lambda_1}t}{\bigg(\frac{3\dot{x}+kx^2+3\sqrt{-\lambda_1}x}
{3\dot{x}+kx^2-3\sqrt{-\lambda_1}x}}\bigg).
\label{lam111}
\end{eqnarray}

We note that unlike the other two examples, equation~(\ref{lam111}) cannot be 
integrated straightforwardly to provide the second
integral (though one can in fact explicitly solve the resultant Riccati equation
after some effort).  We follow the procedure adopted in the previous two 
examples and
construct $I_2$.  Now substituting the expression $S_2$ into (\ref{lam103})   
and solving it in the same way as outlined above we 
obtain the following form for $R$, that is, 
\begin{equation} 
R_{2}=C_0\frac{kx+3\sqrt{-\lambda_1}}
{k(3\dot{x}+kx^2-3\sqrt{-\lambda_1}x)^2}e^{-3\sqrt{\lambda_1}t}.
\label{lam112}
\end{equation}
However, this set $(S_2,R_2)$ does not satisfy the extra constraint 
(\ref{lam104}) and so to deduce the correct form of $R_2$ we assume that 
\begin{align}
\hat{R}_2 & = F(I_1)R_2 = 
C_0\frac{F(I_1)(kx+3\sqrt{-\lambda_1})e^{-3\sqrt{-\lambda_1}t}}
{k(3\dot{x}+kx^2-3\sqrt{-\lambda_1}x)^2}.
\label{lam113}
\end{align}
Substituting (\ref{lam113}) into equation~(\ref{lam104}) we obtain 
$F = {\frac {1}{I_1^2}}$, which fixes the form of $\hat{R}$ as 
\begin{eqnarray}         
\hat{R}_2 = C_0\frac{kx+3\sqrt{-\lambda_1}}
{k(3\dot{x}+kx^2+3\sqrt{-\lambda_1}x)^2}e^{\sqrt{-\lambda_1}t}. 
\label{lam116}  
\end{eqnarray}
Now one can easily check that this set $(S_2,\hat{R}_2)$ is a compatible 
solution for the set (\ref{lam102})-(\ref{lam104}) which in turn provides 
$I_2$ through the relation (\ref{met13}),
\begin{eqnarray} 
I_2=-\frac{2}{k}e^{\sqrt{-\lambda_1}t}
{\bigg(\frac{9\lambda_1+3k\dot{x}+k^2x^2}
{3\dot{x}+kx^2+3\sqrt{-\lambda_1}x}}\bigg).
\label{lam117}
\end{eqnarray}
Using the explicit form of the first integrals $I_1$ and $I_2$, the solution 
can be deduced directly as
\begin{eqnarray}
x=\bigg(\frac{3\sqrt{-\lambda_1}(I_1e^{2\sqrt{-\lambda_1}t}-1)}
{kI_1I_2e^{\sqrt{-\lambda_1}t}+k{(1+I_1e^{2\sqrt{-\lambda_1}t})}}\bigg).
\label{lam119}
\end{eqnarray}
To our knowledge, the above explicit solution, (\ref{lam119}), of the 
equation (\ref{lam101}) is given for the first time.  It has several interesting
consequences for nonlinear dynamics, which will be discussed separately.
\subsection{Type - II Systems}
\label{sec4.2}
In the previous category we considered examples which straightforwardly give 
the integrals $I_1$ and $I_2$ through the relation (\ref{met13}).  In 
the present category we show that there are situations in which an explicit 
form of $I_2$ is difficult to obtain through the relation (\ref{met13}) 
even though one has a compatible solution for (\ref{met9})-(\ref{met11}). 
So one has to 
go for an alternate way in order to obtain the general solution for the 
given problem.  For this purpose, we make use of the method proposed in 
\S3$\,b\,$.
In the following we give examples where such a possibility occurs and how to 
over come this situation.
\subsubsection*{Example 4 : Helmholtz oscillator}
Recently  Almendral \& Sanjuan (2003) studied the invariance and 
integrability properties of the Helmholtz oscillator with friction,
\begin{eqnarray}	   
\ddot{x}+c_1\dot{x}+c_2x-\beta x^2 = 0, 
\label{helm}
\end{eqnarray}
where $c_1$, $c_2$ and $\beta$ are arbitrary parameters, which is a simple 
nonlinear oscillator having a quadratic nonlinearity.  Using the Lie theory
for differential equations Almendral \& Sanjuan (2003) found a  parameteric
choice $c_2 = \frac{6c_1^2}{25}$ for which  the system is integrable and 
derived the general solution for this parametric value.  In the following we 
solve this problem through the extended PS method.  

Substituting $\phi = -(c_1\dot{x}+c_2x-\beta x^2)$ into 
equations~(\ref{met9})-(\ref{met11}) we obtain
\begin{align}            
S_t+\dot{x}S_x-(c_1\dot{x}+c_2x-\beta x^2)S_{\dot{x}} & = c_2
-2\beta x-c_1S+S^2,
\label{cat21}\\
R_t+\dot{x}R_x-(c_1\dot{x}+c_2x-\beta x^2)R_{\dot{x}} & = -R(S-c_1),
\label{cat22}\\
R_x & = SR_{\dot{x}} + RS_{\dot{x}}.
\label{cat23}
\end{align} 
Making the same form of an ansatz, vide equations~(\ref{cat15}) and (\ref{new3}),
we find nontrivial solution exists for (\ref{cat21})-(\ref{cat22}) only
for the parametric restriction $c_2 = \frac{6c_1^2}{25}$.  The respective
solutions are
\begin{align}
S_1 & = \frac {(\frac{2c_1 \dot{x}}{5}+\frac{4c_1^2 x}{25}-{\beta x^2})}
{ \dot{x}+\frac {2c_1}{5}x}, \quad 
R_1  = -( \dot{x}+\frac {2c_1}{5}x)e^{\frac{6}{5}c_1 t},
\label{cat24} \\
S_2 & = \frac {(\frac{c_1 \dot{x}}{5}+\frac{6c_1^2 x}{25}-{\beta x^2})}
{\dot{x}}, \quad R_2  = - \dot{x} e^{c_1t}.
\label{cat25}
\end{align}
Now one can easily check that $(S_1,R_1)$ satisfies the third equation 
(\ref{cat23}) and as a consequence leads to the first integral of the form
\begin{eqnarray}
I_1=e^{\frac{6}{5}c_1 t} \bigg(\frac{\dot{x}^2}{2}+\frac{2c_1 x\dot{x}}{5}
+\frac{2c_1^2 x^2}{25}-\frac{\beta x^3}{3}\bigg).
\label{cat26}
\end{eqnarray}
However, the second set $(S_2,R_2)$ does not satisfy the extra constraint 
(\ref{cat23}) and so we take 
\begin{eqnarray}
\hat{R}_2 = F(I)R_2 = -F(I)\dot{x}e^{c_1t},
\label{cat27}
\end{eqnarray}
which in turn gives $F=C_0{I}^{-\frac{5}{6}}$, where $C_0$ is an integration 
constant, so that 
\begin{eqnarray}          
\hat{R}_2 = -\bigg(\frac{C_0}{I_1^{\frac{5}{6}}}\bigg)\dot{x}e^{c_1t} 
= -\frac {C_0\dot{x}}{({\frac{\dot{x}^2}{2}}+{\frac{2c_1 x\dot{x}}{5}}
+{\frac{2c_1^2 x^2}{25}}-\beta {\frac{x^3}{3}})^{\frac{5}{6}}}.
\label{cat28}
\end{eqnarray} 
One can check that $(S_2,\hat{R}_2)$ satisfy equations~(\ref{cat21})- 
(\ref{cat23}) 
and so one can proceed to deduce the second integration constant through the 
relation (\ref{met13}).  However, upon substituting $(S_2,\hat{R}_2)$ into 
(\ref{met13}) we arrive at 
\begin{eqnarray}
I_2 = \int \frac {c_1\dot{x}+{\frac {6c_1^2x}{25}}-\beta x^2}
{({\frac{\dot{x}^2}{2}}+{\frac{2c_1 x\dot{x}}{5}}
+{\frac{2c_1^2 x^2}{25}}-\beta {\frac{x^3}{3}})^{\frac{5}{6}}}dx. 
\label{cat29}
\end{eqnarray}
It is very difficult to evaluate the integral and so is one is not 
able to obtain an explicit form of $I_2$ for this problem 
through this way.  A similar form of $I_2$ has been also derived by 
Bluman \& Anco (2002) and Jones \textit{et al.} (1993) for the Duffing 
oscillator problem (that is cubic nonlinearity in equation~(\ref{helm})).  

Unlike the other examples discussed in Type I the present example possesses 
difficulties in evaluating the second integration constant.  In fact, 
for a class of equations one faces such complicated integrals.  To overcome 
this one has to look for an alternate way such that the second constant  
can be deduced in a straightforward and simple way.  We tackle this situation 
in the following way.  As we have seen, in most of the problems, we are 
able to deduce the first integral, that is, $I_1$, straightforwardly and 
the first integral often admits explicit time dependent terms.  A useful 
way to overcome this is to remove the explicit time dependent terms by 
transforming the resultant differential equation into an autonomous 
form and integrate the latter and obtain the solution.  In order to do this 
one needs a transformation and the latter can often be constructed through 
ad-hoc way.  However, as we have shown in the theory in \S3$\,b\,$, one can 
deduce the required transformation coordinates in a simple way from the first 
integral itself and the problem can be solved in a systematic way.

Rewriting the first integral $I_1$ given by equation~(\ref{cat26}), in the form 
(\ref{met13a}), we get
\begin{eqnarray}
I_1=\frac{1}{2} \bigg(\dot{x}+\frac{2c_1 x}{5}\bigg)^2 e^{\frac{6}{5}c_1 t}
- \frac{\beta x^3}{3}e^{\frac{6}{5}c_1 t}.
\label{cat200}
\end{eqnarray}
Now splitting the first term in equation~(\ref{cat200}) further in the form 
(\ref{met13b}),
\begin{eqnarray} 
I_1=e^{\frac{2c_1 t}{5}}\bigg(\frac {d}{dt}(\frac{1}{\sqrt{2}}x 
 e^{\frac {2c_1 t}{5}})\bigg)^2-\frac{\beta}{3}(xe^{\frac{2}{5}c_1 t})^3, 
 \label{cat201}
\end{eqnarray}
and identifying the dependent and independent variables from (\ref{cat201}) 
and the relations (\ref{met13c}), we obtain the transformation
\begin{eqnarray} 
w = \frac{1}{\sqrt{2}}x e^{\frac {2c_1 t}{5}}, \quad
z = -\frac {5}{c_1}e^{-\frac {c_1 t}{5}}. 
\label{cat202}
\end{eqnarray}
One can easily check that equation~(\ref{helm}) can be transformed to an autonomous 
form with the help of the transformation (\ref{cat202}).  We note that the
transformation (\ref{cat202}) exactly coincides with the earlier one which has
been constructed via Lie symmetry analysis in Almendral \& Sanjuan (2003).

Using the transformation (\ref{cat202}) the first integral (\ref{cat200}) 
can be rewritten in the form
\begin{equation}
\hat{I} = w'^2-\frac{\hat{\beta}}{3}w^3
\label{newi1}
\end{equation}
which in turn leads to the solution by an integration.  On the other hand the 
transformation changes the equation of motion (\ref{helm}) to  
\begin{eqnarray}
{w''}=\hat{\beta} w^2,
\label{cat203}
\end{eqnarray}
where $\hat{\beta}=2\sqrt{2}\beta$, which upon integration gives 
(\ref{newi1}).  From equation~(\ref{newi1}), we obtain 
\begin{eqnarray}
w'^2=4w^3-g_3,
\label{cat203a}
\end{eqnarray}
where $z=2\sqrt{\frac{3}{\hat{\beta}}}\hat{z}$
and $g_3=-\frac{12I_1}{\hat{\beta}}$. The solution of this differential 
equation can be represented in terms of Weierstrass function 
$\varrho(\hat{z};0,g_3)$ (Gradshteyn \& Ryzhik 1980; Almendral \& Sanjuan 2003).
\subsection{Type - III Systems}
\label{sec4.3}
In the previous two categories we met the situation in which we are able to 
construct a pair of solutions $(S_1,S_2)$ for the equations~(\ref{met9}) 
from which $R_1$ and $R_2$ have been deduced.  However, there are situations 
in which one is able to construct only one set of solution $(R_1,S_1)$ and its
corresponding first integral only and the second pair of solution $(R_2,S_2)$ 
can not be obtained by simple rational form of ansatz.  In this situation 
one can utilize our procedure and deduce the general solution for the given 
problem.  In the following we illustrate this with a couple of examples. 

\subsubsection*{Example 5 : Force free Duffing-van der Pol oscillator}
One of the well-studied but still challenging equations in nonlinear dynamics
is the Duffing-van der Pol oscillator equation. Its 
autonomous version 
(force-free) is
\begin{eqnarray} 
\ddot{x}+(\alpha+\beta x^2)\dot{x}-\gamma x +x^3=0, 
\label{dvp}
\end{eqnarray}
where  over dot denotes differentiation with respect to time and 
$\alpha$, $\beta$ and $ \gamma$ are arbitrary parameters. Equation (\ref{dvp})
arises in a model describing the propagation of voltage pulses along a 
neuronal axon and has received a lot of attention recently by many authors.  
A vast amount of literature exists on this equation, for details see for 
example Lakshmanan \& Rajasekar (2003) and references therein.  In this case 
we have
\begin{align}            
S_t+\dot{x}S_x-((\alpha+\beta x^2)\dot{x}-\gamma x +x^3)S_{\dot{x}}
& = (2\beta x\dot{x}-\gamma  +3x^2)
\nonumber\\
& \quad -(\alpha+\beta x^2)S+S^2, 
\label{cat301}\\
R_t+\dot{x}R_x-((\alpha+\beta x^2)\dot{x}-\gamma x +x^3)R_{\dot{x}}
& = (\alpha+\beta x^2-S)R, 
\label{cat302}\\
R_x & = SR_{\dot{x}}+RS_{\dot{x}}. 
\label{cat303}
\end{align} 
To solve equation~(\ref{cat301})-(\ref{cat303}) we seek an ansatz for 
 $S$ and $R$ of the form 
\begin{eqnarray}
S = \frac{a(t,x)+b(t,x)\dot{x}}{c(t,x)+d(t,x)\dot{x}}, \quad
R = A(t,x)+B(t,x)\dot{x}.
\label{cat304}
\end{eqnarray}
Upon solving the equations~(\ref{cat301})-(\ref{cat303}) with the above ansatz 
we find nontrivial solution exists only for the choice 
$\alpha=\frac {4}{\beta}, \gamma=-\frac{3}{\beta^2}$ and the corresponding 
forms of $S$ and $R$ reads 
\begin{eqnarray}
S = \frac {1}{\beta}+\beta x^2, \quad
R = e^{\frac {3t}{\beta}}.
\label{cat305}
\end{eqnarray}
For this set one can construct an invariant through the expression 
(\ref{met13}) which turns out to be (Senthilvelan \& Lakshmanan 1995) 
\begin{eqnarray} 
\dot{x}+\frac{1}{\beta}x+\frac{\beta}{3}x^3=Ie^{-\frac{3}{\beta}t}. 
\label{dvp1}
\end{eqnarray}
To obtain a second pair of solutions for the equations~(\ref{cat301})-(\ref{cat303})
one may seek more general rational form of $S$ and $R$ by including higher
polynomials in $\dot{x}$.  However, they all lead to only functionally dependent
integrals.  As it is not possible to seek the second pair of solution by simple
ansatz one has to see an alternate way as indicated in \S3$\,b\,$.  We can 
deduce the required transformation coordinates from the first integral and  
transform the latter to an autonomous equation and integrate it.  

Using our algorithm given in \S3$\,b\,$ one can deduce the transformation
coordinates from the first integral itself which turns out to be
(Chandrasekar \textit{et al.} 2004)
\begin{eqnarray} 
w= -x e^{\frac{1}{\beta}t},\;\;\;\;\; 
z= e^{-\frac{2}{\beta}t},\label{cat311}
\end{eqnarray}
where $w$ and $z$ are new dependent and independent variables respectively.  
Substituting (\ref{cat311}) into (\ref{dvp}) with the parametric restriction
$\alpha=\frac {4}{\beta}, \gamma=-\frac{3}{\beta^2}$, we get 
\begin{eqnarray} 
w''-\frac {\beta^2}{2}w^2w'=0,
\label{cat312}
\end{eqnarray}
where prime denotes differentiation with respect to $z$.  Equation 
(\ref{cat312}) can be integrated trivially to yield 
\begin{eqnarray}
w'-\frac{\beta^2}{6}w^3=I, 
\label{cat313}
\end{eqnarray}
where $I$ is the integration constant.  Equivalently, the transformation 
(\ref{cat311}) reduces (\ref{dvp}) to this form.  Solving (\ref{cat313}),
we obtain (Gradshteyn \& Ryzhik 1980) 
\begin{eqnarray} 
z-z_0=\frac {a}{3I}\left[\frac{1}{2}
\log\left(\frac{(w+a)^2}{w^2-aw+a^2}\right)
+\sqrt{3}\,\mbox{arctan}\left(\frac{w\sqrt{3}}{2a-w}\right)\right], 
\label{cat314}
\end{eqnarray}
where $a=\sqrt[3]{\frac {6I}{\beta ^2}}$
and $z_0$ is the second integration constant.  Rewriting $w$ and $z$ in terms 
of old variables one can get the explicit solution for the equation 
(\ref{dvp}). 

In the above, we have shown that the systems (\ref{helm}) and (\ref{dvp}) are 
integrable for certain
specific parametric restrictions only. One may also assume that the functions
$S$ and $R$ involve higher degree rational functions in $\dot{x}$ and repeat the
analysis. However, such an analysis does not provide any new integrable choice.
In fact, the present results
coincide exactly with the results obtained through other methods, namely, 
Painlev\'e analysis, Lie symmetry analysis and direct methods 
(Senthilvelan \& Lakshmanan 1995; Almendral \& Sanjuan 2003; 
Lakshmanan \& Rajasekar 2003).
\section{Linearizable equations}
\label{sec5}
In the previous section we discussed the complete integrability of 
nonlinear dynamical systems by constructing sufficient number of integrals of 
motion and obtaining the general solutions explicitly.  Another way of
solving nonlinear ODEs is to transform them to linear ODEs, 
in particular to a free particle equation and explore their underlying
solutions.  Eventhough this is one of the classical problems in the theory of 
ODEs, recently considerable progress has been made 
(Mahomed \& Leach 1989$a$; Steeb 1993; Olver 1995; Harrison 2002).  In this 
direction it has 
been shown that a necessary condition for a second order ODE to be 
linearizable is that it should be of the form (Mahomed \& Leach 1989$a$)
\begin{eqnarray}
\ddot{q}=D(t,q)+C(t,q)\dot{q}+B(t,q)\dot{q}^2+A(t,q)\dot{q}^3,
\label {lt01} 
\end{eqnarray}
where the functions $A$, $B$, $C,$ and $D$ are analytic.  Sufficient condition 
for the
above second order equation to be linearizable is (Mahomed \& Leach 1989$a$),
\begin{eqnarray}
3A_{tt}+3CA_{t}-3DA_{q}+3AC_{t}+C_{qq}-6AD_{q}+BC_{q}-2BB_{t}-2B_{tq}=0,
\nonumber\\
B_{tt}+6DA_{t}-3DB_{q}+3AD_{t}-2C_{tq}-3BD_{q}+3D_{qq}+2CC_{q}-CB_{t}=0,
\label {lt02} 
\end{eqnarray}
where the suffices refer to partial derivatives.

For a given second order nonlinear ODE one can easily check whether it can be
linearizable or not by using the above necessary and sufficient conditions.
However, the nontrivial problem is how to deduce systematically the
linearizing transformations if the given equation is linearizable.  As far our
knowledge goes Lie symmetries are often used to extract the linearizing
transformations (Mahomed \& Leach 1985).  
As we pointed out in \S3 the linearizing
transformations can also be deduced from the first integral itself, whenever
the system is linearizable, in a simple and straightforward way and we stress
that our procedure
is new to the literature.  In fact, we use the
same procedure discussed in \S3$\,c\,$ and deduce the linearizing transformations.
The only difference is that in the case of linearizing transformations the
function $F_2$ turns out to be zero in equation~(\ref{met13b}) and as a 
consequence
the latter becomes $\frac {dw}{dz}=I$ and the transformation
coordinates become the linearizing transformations.  We illustrate
the theory with certain new examples in the following.

\subsubsection*{Example 1 : General relativity}
To illustrate the underlying ideas let us begin with a simple and physically
interesting example, namely, the general relativity equation which we
discussed as Example 1 in \S4.  We derived the solution (\ref{app38})
using the PS method.  In the present section we linearize the system and derive
its solution.  Rewriting the first integral (\ref{cat16}) in the form 
(\ref{met13a})
\begin{eqnarray}
I=-\frac{1}{2t}\frac {d}{dt}\left(\frac{1}{x^2}\right),
\label {gr01} 
\end{eqnarray}
and identifying (\ref{gr01}) with (\ref{met13b}), we get
\begin{eqnarray}
G_1=\frac{1}{x^2},\;\;\;\;\; 
G_2=-2t,\;\;\;\;\;  F_2=0.
\label {gr02} 
\end{eqnarray}
With the above choices, equation~(\ref{met13c}) furnishes the transformed
variables,
\begin{eqnarray}
w=\frac{1}{x^2},\;\;\;\;\;  
z=-t^2.
\label {gr03}  
\end{eqnarray}
Substituting (\ref{gr03}) into (\ref{cat11}), the latter becomes the free
particle equation, namely, $\frac{d^2w}{dz^2}=0$, whose general solution is 
$w=I_1z+I_2$, where $I_1$ and $I_2$ are integration constants.  Rewriting 
$w$ and $z$ in terms of $x$ and $t$ one gets 
exactly (\ref{app38}) which has been derived in a different way.

\subsubsection*{Example 2 : Modified Emden type equations }
Recently several papers have been devoted to explore the invariance and
integrable properties of the modified Emden type equations 
(Mahomed \& Leach 1985; Duarte \textit{et al.} 1987),
\begin{eqnarray}            
\ddot{x}+kx\dot{x}+\frac {k^2}{9}x^3=0.
\label {lin101a}
\end{eqnarray}
In fact, it is one of the rare second order nonlinear ODEs  
which admit eight Lie point symmetries and
as a consequence is a linearizable one.  Recently Pandey \textit{et al.} 
(2004) have 
obtained the explicit forms of the Lie point symmetries associated with the 
more general equation
\begin{eqnarray}            
\ddot{x}+kx\dot{x}+\frac {k^2}{9}x^3+\lambda_1x+\lambda_2=0,
\label {lin101}
\end{eqnarray}
where $k$, $\lambda_1$ and $\lambda_2$ are arbitrary parameters. They found 
that not only the Emden equation (\ref{lin101a}), but 
also its general form, that is, equation~(\ref{lin101}), admits eight 
Lie point symmetries.  The authors have also reported that
the explicit forms of the symmetry generators.  However, due to the complicated
forms of the symmetry generators it is difficult to derive the first integrals
and linearizing transformations from the symmetries straightforwardly 
(though in principle this is always possible).  Nevertheless, we discussed
about the integrability of the case $\lambda_2=0$, $\lambda_1\neq0$ of 
equation~(\ref{lin101}) as Example 3 in \S4 and deduced its general solution.  
In this section we 
transform the equation into free particle equation and deduce the general
solution in an independent way.  We divide our analysis into 2 cases,
namely, (i) $\lambda_1\neq0$, $\lambda_2=0$ and
(ii) $\lambda_1\neq0$, $\lambda_2\neq0$ and construct linearizing 
transformations and general solutions for both the cases.  As the
procedure is same as given in the previous examples we give only the results.

\subsubsection*{Case(i) $\lambda_2=0$, $\lambda_1\neq0$ :  
Modified Emden type equation with linear term}
Restricting $\lambda_2=0$ in (\ref{lin101}) we have
\begin{eqnarray}            
\ddot{x}+kx\dot{x}+\frac {k^2}{9}x^3+\lambda_1x=0.
\label {lam102a}
\end{eqnarray}
Since the first integral is already derived, vide equation~(\ref{lam111}), 
we utilize it here to deduce the linearizing transformations.  Rewriting the 
first integral (\ref{lam111}) in the form
\begin{eqnarray}
I_1=-\frac{e^{-\sqrt{-\lambda_1}t}kx^2}
{{3\dot{x}+kx^2-3\sqrt{-\lambda_1}x}}\left[\frac {d}{dt}
\left((\frac{3}{kx}+\frac{1}{\sqrt{-\lambda_1}})e^{-\sqrt{-\lambda_1}t}
\right)\right]
\label {lin103} 
\end{eqnarray}
and identifying (\ref{lin103}) with (\ref{met13b}), we get
\begin{eqnarray}
G_1  = \bigg(\frac{3}{kx}+\frac{1}{\sqrt{-\lambda_1}}\bigg)
e^{-\sqrt{-\lambda_1}t},\quad
G_2  = -\frac{{3\dot{x}+kx^2-3\sqrt{-\lambda_1}x}}{kx^2}
e^{\sqrt{-\lambda_1}t}.
\end{eqnarray}
With the above functions (\ref{met13c}) furnishes
\begin{eqnarray}
w  = \bigg(\frac{3}{kx}+\frac{1}{\sqrt{-\lambda_1}}\bigg)
e^{-\sqrt{-\lambda_1}t},\quad 
z  = \bigg(\frac{3}{kx}-\frac{1}{\sqrt{-\lambda_1}}\bigg)
e^{\sqrt{-\lambda_1}t},
\label {lin104}
\end{eqnarray}
which is nothing but the linearizing transformation. One may note that in this
case also while rewriting the first integral $I$ (equation~(\ref{lam111})) in 
the form (\ref{met13a}), the function  $F_2$ disappears, and as a 
consequence we arrive at (vide equation~(\ref{met13d})) 
\begin{eqnarray} 
\frac {dw}{dz}=I,
\end{eqnarray}
which in turn gives the free particle equation by differentiation or leads to 
the solution (\ref{lam119}) by an integration.  On the other hand vanishing of 
the function $F_2$ in this analysis is precisely the condition for the system 
to be transformed into the free particle equation.

\subsubsection*{Case(ii) $\lambda_1\neq0$, $\lambda_2\neq0$ : Modified Emden 
type equation with linear term and constant external forcing}
Finally, we consider the general case, that is,
\begin{eqnarray}            
\ddot{x}+kx\dot{x}+\frac {k^2}{9}x^3+\lambda_1x+\lambda_2=0.
\label {lin302}
\end{eqnarray}
To explore the first integrals associated with the system (\ref{lin302})
let us seek the PS algorithm again.  The determining equations for the functions
$R$ and $S$ turn out to be
\begin{align}           
S_t+\dot{x}S_x-(kx\dot{x}+\frac {k^2}{9}x^3+\lambda_1x+\lambda_2) 
S_{\dot{x}} & = 
k\dot{x}+\frac {k^2}{3}x^2+\lambda_1-S kx+S^2,\label{lin303}\\
R_t+\dot{x}R_x-(kx\dot{x}+\frac {k^2}{9}x^3+\lambda_1x+\lambda_2)
R_{\dot{x}} & = (kx-S)R,\label {lin304}\\
R_x-SR_{\dot{x}}-RS_{\dot{x}} & = 0.\label {lin305}
\end{align}
As before let us seek an ansatz for $S$ to solve the equation~(\ref{lin303}), 
namely, 
\begin{eqnarray}
S = \frac{a(t,x)+b(t,x)\dot{x}}{c(t,x)+d(t,x)\dot{x}}.
\label{lin305a}
\end{eqnarray}
Substituting (\ref{lin305a}) into (\ref{lin303}) and equating the coefficients 
of different powers of $\dot{x}$ to zero and solving the resultant equations we
arrive at 
\begin{equation}
S_1=\frac{kx+3\alpha}{3}-\frac{k\dot{x}}
{kx+3\alpha}, \quad 
S_2=\frac{kx+3\beta }{3}-\frac{k\dot{x}}
{kx+3\beta},
\label{lin306}
\end{equation}
where $\alpha^3+\alpha\lambda_1-\frac{k\lambda_2}{3}=0$ and $\beta=
\frac{-\alpha\pm\sqrt{-3\alpha^2-4\lambda_1}}{2}$.  Putting the forms of $S_1$ 
 into (\ref{lin304}) we get
\begin{equation}
R_t+\dot{x}R_x-(kx\dot{x}+\frac {k^2}{9}x^3+\lambda_1x+\lambda_2) 
R_{\dot{x}}=\bigg(\frac{k\dot{x}}{kx+3\alpha}-\frac{kx+3\alpha}{3}+
kx\bigg)R.
\label{lin307}
\end{equation}
Again to solve this equation we make an ansatz
\begin{equation}
R = \frac{A(t,x)+B(t,x)\dot{x}}{C(t,x)+D(t,x)\dot{x}+E(t,x)\dot{x}^2}.
\label{lin307a}
\end{equation} 
Substituting (\ref{lin307a}) into (\ref{lin307}) and solving it we obtain
the following form of $R$, namely,
\begin{equation}
R_1=\frac{C_0(kx+3\alpha)e^{\mp\hat{\alpha}t}}
{(3k\dot{x}-3\frac{(3\alpha\pm\hat{\alpha})}{2}
(kx+3\alpha)+(kx+3\alpha)^2)^2},
\label{lin308}
\end{equation}
where $C_0$ is constant and $\hat{\alpha}=\sqrt{-3\alpha^2-4\lambda_1}$.
We find that the solution ($S_1$,$R_1$) satisfies (\ref{lin305}).  
Equations~(\ref{lin306}) and (\ref{lin308}) fix the first integral of the form
\begin{eqnarray}
I_1=e^{\mp\hat{\alpha}t}
\bigg(\frac{3k\dot{x}-3\frac{(3\alpha\mp\hat{\alpha})}
{2}(kx+3\alpha)+(kx+3\alpha)^2}
{3k\dot{x}-3\frac{(3\alpha\pm\hat{\alpha})}{2}
(kx+3\alpha)+(kx+3\alpha)^2}\bigg),
\label{lin309}
\end{eqnarray}
where $C_0=9k\hat{\alpha}$.  Rewriting the first integral (\ref{lin309}) in 
the form (\ref{met13a})
\begin{eqnarray}
I_1=-\frac{e^{\frac{-3\alpha\mp\hat{\alpha}}{2}t}(k_1x+3\alpha)^2}
{3k\dot{x}-3\frac{(3\alpha\pm\hat{\alpha})}{2}
 (kx+3\alpha)+(kx+3\alpha)^2}\nonumber\\
 \times\left[\frac {d}{dt}
\left((\frac{-3}{kx+3\alpha}+\frac{3\alpha\pm\hat{\alpha}}
{2(3\alpha^2+\lambda_1)})
e^{\frac{3\alpha\mp\hat{\alpha}}{2}t}\right)\right]
\label {lin310} 
\end{eqnarray}
and identifying (\ref{lin310}) with (\ref{met13b}), we get
\begin{align}
G_1 & = \bigg(\frac{-3}{kx+3\alpha}+\frac{3\alpha\pm\hat{\alpha}}
{2(3\alpha^2+\lambda_1)}\bigg)
e^{\frac{3\alpha\mp\hat{\alpha}}{2}t},\nonumber\\
G_2 & = -\frac{3k\dot{x}-3\frac{(3\alpha\pm\hat{\alpha})}{2}
 (kx+3\alpha)+(kx+3\alpha)^2}{(kx+3\alpha)^2}
 e^{\frac{3\alpha\pm\hat{\alpha}}{2}t},
\end{align}
so that (\ref{met13c}) gives
\begin{align}
w & = \bigg(\frac{-3}{kx+3\alpha}+\frac{3\alpha\pm\hat{\alpha}}
{2(3\alpha^2+\lambda_1)}\bigg)
e^{\frac{3\alpha\mp\hat{\alpha}}{2}t},\nonumber\\
z & = \bigg(\frac{-3}{kx+3\alpha}+\frac{3\alpha\mp
\hat{\alpha}}{2(3\alpha^2+\lambda_1)}\bigg)e^{\frac{3\alpha
\pm\hat{\alpha}}{2}t},
\label {lin311}
\end{align}
which is nothing but the linearizing transformation.  Substituting 
(\ref{lin311}) into (\ref{lin302}) we get the free particle equation
\begin{eqnarray} 
\frac {d^2w}{dz^2}=0
\label {lin311a}
\end{eqnarray}
whose general can be written as $w=I_1z+I_2$.  Rewriting $w$ and $z$ in 
terms of the original variable $x$ and $t$ one obtains
\begin{eqnarray}
x=-\frac{3\alpha}{k}+\frac{6}{k}
{\bigg(\frac{(3\alpha^2+\lambda_1)(1-I_1e^{\pm\hat{\alpha}t})}
{3\alpha(1-I_1e^{\pm\hat{\alpha}t})
-2(3\alpha^2+\lambda_1)I_2
e^{\frac{-3\alpha\pm\hat{\alpha}}{2}t}
\pm \hat{\alpha}{(1+I_1e^{\pm\hat{\alpha}t})}}
\bigg)}.
\label{lin312}
\end{eqnarray}
  
On the other hand the general solution can also be derived by extending the PS 
method itself.  To do so one has to consider the function $S_2$.
Now substituting the expression $S_2$ into (\ref{lin304}) 
and solving it in the same way as outlined in the previous paragraphs we 
obtain the following form for $R$, that is,
\begin{equation} 
R_{2}=\frac{C_0(kx+3\beta)e^{\frac{3(\alpha\mp\hat{\alpha})}{2}t}}
{(3k\dot{x}-3\frac{(3\alpha\pm\hat{\alpha})}{2}
(kx+3\alpha)+(kx+3\alpha)^2)^2}.
\label{lin312a}
\end{equation}
However, this set, $(S_2,R_2)$, does not satisfy the extra constraint 
(\ref{lin305}) and to recover the full form of the integrating factor we 
assume that 
\begin{align}
\hat{R}_2 & = F(I_1)R_2.
\label{lin312b}
\end{align}
Substituting (\ref{lin312b}) into equation~(\ref{lin305}) we obtain an equation 
for $F$, that is, $I_1F'+2F = 0$, where prime denotes differentiation with 
respect to $I_1$.  Upon integrating this equation we obtain 
$F = 1/I_1^2$, which fixes the form of $\hat{R}$ as  
\begin{eqnarray}         
\hat{R}_2 =\frac{C_0(kx+3\beta)e^{\frac{3\alpha\pm\hat{\alpha}}{2}t}}
{(3k\dot{x}-3\frac{(3\alpha\mp\hat{\alpha})}{2}
(kx+3\alpha)+(kx+3\alpha)^2)^2}. 
\label{lin313}  
\end{eqnarray}
Now one can easily check that this set $S_2$ and $\hat{R}_2$ is a compatible 
solution for the equations~(\ref{lin303})-(\ref{lin305}).  Substituting $S_2$ 
and $\hat{R}_2$ into (\ref{met13}), we can obtain an explicit form for the 
second integral $I_2$, that is,
\begin{eqnarray} 
I_2=-\bigg(\frac{2\hat{\alpha}(3k\dot{x}-3\alpha kx+
k^2x^2+9\alpha^2+9\lambda_1)e^{\frac{3\alpha\pm\hat{\alpha}}{2}t}}
{(3\alpha\pm\hat{\alpha})(3k\dot{x}-3\frac{(3\alpha\mp\hat{\alpha})}{2}
(kx+3\alpha)+(kx+3\alpha)^2)}\bigg).
\label{lin314}
\end{eqnarray}
Rewriting equation~(\ref{lin309}) for $\dot{x}$ and substituting it into 
(\ref{lin314}) we get the same expression (\ref{lin312}) as the 
general solution.

\subsubsection*{Example 3 : Generalized modified Emden type equation}
Recently, Pandey \textit{et al.} (2004)  have considered the following Li\'enard 
equation
\begin{eqnarray}            
\ddot{x}+f(x)\dot{x}+g(x)=0,
\label {lin01a}
\end{eqnarray}
where $f$ and $g$ are arbitrary function of their arguments, and classified 
systematically all polynomial forms of $f$ and $g$ which admit  
eight Lie point symmetry 
generators with their explicit forms.  They found that the most general 
nonlinear ODE which is linear in 
$\dot{x}$ whose coefficients are functions of the dependent variable alone 
should be of the form
\begin{eqnarray}            
\ddot{x}+(k_1x+k_2)\dot{x}+\frac {{k_1}^2}{9}x^3+\frac{k_1k_2}{3}x^2+
\lambda_1x+\lambda_2=0,
\label {lin01}
\end{eqnarray}
where $k_i$'s and $\lambda_i$'s, i=1,2, are arbitrary parameters, which is
consistent with the criteria (\ref{lt01}) and (\ref{lt02}) given by Mahomed 
\& Leach (1989$a$).  Interestingly,
equation~(\ref{lin01}) and all its sub cases posses $sl(3,R)$ symmetry 
algebra.  For example,
we discussed the integrability and linearization of equation~(\ref{lin01}) with
$k_2=0$ in the previous example.  As the linearizing transformations and the
general solution of equation~(\ref{lin01}) are yet to be reported we include
this equation as an example in the present work.
As in the previous case we divide our analysis into three cases.\\
(i) $\lambda_1=0$, $\lambda_2=0$ : Modified Emden type equation with 
quadratic and cubic nonlinearity
\begin{eqnarray}            
\ddot{x}+(k_1x+k_2)\dot{x}+\frac {{k_1}^2}{9}x^3+\frac{k_1k_2}{3}x^2=0.
\label {lin02}
\end{eqnarray}
(ii)$\lambda_1\neq0$, $\lambda_2=0$ : Modified Emden type equation 
with quadratic and linear terms
\begin{eqnarray}            
\ddot{x}+(k_1x+k_2)\dot{x}+\frac {{k_1}^2}{9}x^3+\frac{k_1k_2}{3}x^2
+\lambda_1x=0.\label {lin03}
\end{eqnarray}
(iii)$\lambda_1\neq0$, $\lambda_2\neq0$ : The full generalized modified Emden 
type equation~(\ref{lin01}).  

We have derived the integrating factors, integrals of motion, linearizing 
transformations and the general solutions for all the cases.  As the 
calculations are the similar
to the one discussed in the previous case, we present the results in tabular 
form (Table I), where the results for the most general case (\ref{lin01}) has
been given from which the results for the limiting cases (\ref{lin02}) and
(\ref{lin03}) can be deduced. 
\section{Conclusion}
\label{sec6}
	In this paper we have discussed the method of finding general 
solutions
associated with second order nonlinear ODEs through a modified PS method.  The
method can be considered as a direct one, complimenting the well known method of
Lie symmetries.  In particular, we have extended
the theory of Duarte \textit{et al.} (2001) such that one can recover new integrating
factors and their associated integrals of motion.  These integrals of motion can
be utilized to construct the general solution.  In the
situation when one is not able to recover the second integral of motion we
introduced another approach to derive the second integration constant.
Interestingly, we showed that in this case it can be derived from the first 
integral itself
in a simple and elegant way.  Apart from the above we introduced a technique
which can be utilized to derive linearizing transformation from the first
integral.  We illustrated the theory with several new examples and explored
their underlying solutions.

In this paper we concentrated our studies only on single second order
ODEs.  In principle the method can also be extended to third order ODEs and
systems of second order ODEs.  The results will be published elsewhere.

The work of VKC is supported by CSIR in the form of a Junior Research
Fellowship.  The work of MS and ML forms part of a Department of Science 
and Technology, Government of India, sponsored research project.

\begin{table}
\caption{Integrating factors, integrals of motion, linearizing transformations
and the general solution of equation~(\ref{lin01})}
\begin{tabular}{ll}
\hline
\multicolumn{2}{l}{Null forms and Integrating factors} \\ \\
$S_1 = \displaystyle \frac{k_1x+3\alpha}{3}-\frac{k_1\dot{x}}{k_1x+3\alpha}$,&
$R_1 =\displaystyle\frac{C_0(k_1x+3\alpha)e^{\mp\hat{\alpha}t}}
{(3k_1\dot{x}-\frac{\hat{\beta}\pm\hat{\alpha}}{2}
(3k_1x+9\alpha)+(k_1x+3\alpha)^2)^2} $\\ \\
$S_2 = \displaystyle  \frac{k_1x+3\beta }{3}-\frac{k_1\dot{x}}{k_1x+3\beta }$,& 
$R_2 =\displaystyle \frac{C_0(k_1x+3\beta)e^{\frac{\hat{\beta}\pm\hat{\alpha}}{2}t}}
{3k_1\dot{x}-\frac{\hat{\beta}\mp\hat{\alpha}}
{2}(3k_1x+9\alpha)+(k_1x+3\alpha)^2} $ \\ \\
$\displaystyle  \alpha^3-k_2\alpha^2+\alpha\lambda_1-\frac{k_1\lambda_2}{3}=0$,
&
$\hat{\alpha}=\sqrt{-3\alpha^2+2\alpha k_2+k_2^2-4\lambda_1}$,
$\hat{\beta}=3\alpha-k_2$ \\ 
$\beta= \displaystyle \frac{-\alpha+k_2\pm\hat{\alpha}}{2}$,& \\ \\
\hline
\multicolumn{2}{l}{First Integrals} \\\\
\multicolumn{2}{l}{$I_1 =\displaystyle  e^{\mp\hat{\alpha}t}
\bigg(\frac{3k_1\dot{x}-\frac{\hat{\beta}\mp\hat{\alpha}}{2}(3k_1x+9\alpha)+
(k_1x+3\alpha)^2}{3k_1\dot{x}-\frac{\hat{\beta}\pm\hat{\alpha}}{2}
(3k_1x+9\alpha)+(k_1x+3\alpha)^2}\bigg)$,\;\;
$C_0=9k_1\hat{\alpha}$}\\ \\

\multicolumn{2}{l}{$I_2 =\displaystyle  \frac{-2\hat{\alpha}
e^{\frac{\hat{\beta}\pm\hat{\alpha}}{2}t}}{\hat{\beta}\pm\hat{\alpha}}
{\bigg(\frac{3k_1\dot{x}-3k_1x(\alpha-k_2)
+k_1^2x^2+9\alpha^2-9\alpha k_2+9\lambda_1}
{3k_1\dot{x}-\frac{\hat{\beta}\mp\hat{\alpha}}{2}
(3k_1x+9\alpha)+(k_1x+3\alpha)^2}\bigg)} $} \\\\
 
\hline
\multicolumn{2}{l}{Linearizing Transformations} \\\\
\multicolumn{2}{l}{ $w = \displaystyle \bigg(\frac{-3}{k_1x+3\alpha}+\frac{\hat{\beta}\pm\hat{\alpha}}
{2(3\alpha^2-2\alpha k_2+\lambda_1)}\bigg)
e^{\frac{\hat{\beta}\mp\hat{\alpha}}{2}t}$,} \\ \\
\multicolumn{2}{l}{ $z =\displaystyle \bigg(\frac{-3}{k_1x+3\alpha}+\frac{\hat{\beta}\mp\hat{\alpha}}
{2(3\alpha^2-2\alpha k_2+\lambda_1)}\bigg)
e^{\frac{\hat{\beta}\pm\hat{\alpha}}{2}t} $ 
}
\\ \\
\hline
\multicolumn{2}{l}{Solution} \\ \\
\multicolumn{2}{l}{$x =\displaystyle -\frac{3\alpha}{k_1}+\frac{1}{k_1}
{\bigg(\frac{6(3\alpha^2-2\alpha k_2+\lambda_1)(1-I_1e^{\pm\hat{\alpha}t})}
{\hat{\beta}(1-I_1e^{\pm\hat{\alpha}t})
\pm (\hat{\beta}\pm\hat{\alpha})I_1I_2e^{\frac{-\hat{\beta}\pm\hat{\alpha}}{2}t}
\pm \hat{\alpha}{(1+I_1e^{\pm\hat{\alpha}t})}}\bigg)} $ } \\\\
\hline
\end{tabular}
\end{table}

\label{lastpage}
\end{document}